\documentclass[a4paper,fleqn,usenatbib]{mnras}
\usepackage[T1]{fontenc}
\usepackage{ae,aecompl}
\usepackage{graphicx}	% Including figure files
\usepackage{amsmath}	% Advanced maths commands
\usepackage{amssymb}	% Extra maths symbols
\usepackage{newtxtext,newtxmath}
\DeclareMathOperator{\sech}{sech}

\title[Fast wave trains: from tadpoles to boomerangs]{Fast magnetoacoustic wave trains: from tadpoles to boomerangs}
\author[D. Y. Kolotkov et al.]{
Dmitrii Y. Kolotkov$^{1,2}$\thanks{E-mail: D.Kolotkov.1@warwick.ac.uk (DYK)},
Valery M. Nakariakov$^{1,3}$,
Guy Moss$^{1}$,
Paul Shellard$^{1}$
\\
$^{1}$Centre for Fusion, Space and Astrophysics, Physics Department, University of Warwick, Coventry CV4 7AL, United Kingdom\\
$^{2}$Institute of Solar-Terrestrial Physics SB RAS, Irkutsk 664033, Russia\\
$^{3}$St. Petersburg Branch, Special Astrophysical Observatory, Russian Academy of Sciences, 196140, St. Petersburg, Russia
}

\date{Accepted 2021 May 27. Received 2021 May 26; in original form 2021 April 19}

\pubyear{2021}

\begin{document}
\label{firstpage}
\pagerange{\pageref{firstpage}--\pageref{lastpage}}
\maketitle

\begin{abstract}
Rapidly propagating fast magnetoacoustic wave trains guided by field-aligned plasma non-uniformities are confidently observed in the Sun's corona. Observations at large heights suggest that fast wave trains can travel long distances from the excitation locations.
We study characteristic time signatures of fully developed, dispersive fast magnetoacoustic wave trains in field-aligned zero-$\beta$ plasma slabs in the linear regime. Fast wave trains are excited by a spatially localised impulsive driver and propagate along the waveguide as prescribed by the waveguide-caused dispersion.
In slabs with steeper transverse density profiles, developed wave trains are shown to consist of three distinct phases: a long-period quasi-periodic phase with the oscillation period shortening with time, a multi-periodic (peloton) phase in which distinctly different periods co-exist, and a short-lived periodic Airy phase. The appearance of these phases is attributed to a non-monotonic dependence of the fast wave group speed on the parallel wavenumber due to the waveguide dispersion, and is shown to be different for axisymmetric (sausage) and non-axisymmetric (kink) modes. In wavelet analysis, this corresponds to the transition from the previously known tadpole shape to a new boomerang shape of the wave train spectrum, with two well-pronounced arms at shorter and longer periods.
We describe a specific previously published radio observation of a coronal fast wave train, {highly suggestive of a change of the wavelet spectrum from a tadpole to a boomerang, broadly consistent with our modelling.} The applicability of these boomerang-shaped fast wave trains for probing the transverse structuring of the waveguiding coronal plasma is discussed.
\end{abstract}


%%%%%%%%%%%%%%%%%%%%%%%%%%%%%%%%%%%%%%%%%%%%%%%%%%

%%%%%%%%%%%%%%%%% BODY OF PAPER %%%%%%%%%%%%%%%%%%

\section{Introduction} \label{sec:intro}

The highly filamented nature of the plasma of the solar corona plays a crucial role in magnetohydrodynamic (MHD) wave processes ubiquitously observed in the corona with spaceborne and ground-based instruments in multiple bands \citep[e.g.,][]{2000SoPh..193..139R}. Various coronal plasma structures filamented along a guiding magnetic field act as effective waveguides for MHD waves, whose dynamics is thereby very much sensitive to the parameters of the host structure. This close connection between the observed behaviour of MHD waves with local coronal plasma conditions creates a solid ground for the plasma diagnostics by the method of coronal MHD seismology. The method is based on the interplay of a theoretical modelling and observations of coronal MHD waves \citep[see e.g.,][for comprehensive reviews]{2012RSPTA.370.3193D, 2016SSRv..200...75N, 2016GMS...216..395W, 2020ARA&A..58..441N}.

Among the phenomena associated with essentially compressive fast magnetoacoustic waves guided by coronal plasma non-uniformities are rapidly propagating quasi-periodic wave trains, observed in various electromagnetic bands, from Extreme Ultraviolet (EUV) and visible light to metric radio waves \citep[see the most recent review by][partly addressing this issue]{2020SSRv..216..136L}. The first direct observation of fast wave trains in the corona was made during a total solar eclipse \citep[see e.g.,][]{2001MNRAS.326..428W, 2003A&A...406..709K}, while the periodic variations of the polarised brightness, discovered by \citet[][]{1997ApJ...491L.111O}, could also be associated with fast waves. The launch of the Solar Dynamics Observatory (SDO) with the instrument Atmospheric Imaging Assembly (AIA/SDO) onboard allowed for the direct observation of fast wave trains in EUV, in the form of rapidly propagating quasi-periodic EUV intensity disturbances localised in space and time in the lower corona \citep{2011ApJ...736L..13L}, followed by a number of other EUV detections \citep[e.g.,][]{2012ApJ...753...52L, 2012ApJ...753...53S, 2013A&A...554A.144Y, 2013SoPh..288..585S, 2014A&A...569A..12N, 2015A&A...581A..78Z, 2018ApJ...853....1S, 2018ApJ...860...54O}. In the observational papers, this phenomenon is often referred to as \lq\lq quasi-periodic fast-propagating\rq\rq\ waves, or QFP waves. Despite the lack of comprehensive statistics of EUV observations of fast wave trains, the majority of observational works report rather similar properties, i.e. the apparent propagation speed from several hundred to a few thousand km\,s$^{-1}$ (about the expected local Alfv\'en speed), oscillation periods between one and a few minutes, and amplitudes up to several percent. 

In the solar radio bursts associated with the plasma emission mechanism, fast wave trains are indirectly observed through the modulation of the local plasma density leading to the appearance of quasi-periodic patterns and fine structuring in the dynamic spectrum of the burst. The high temporal resolution of radio observations allowed for revealing fast wave trains with second \citep[e.g.,][]{2013A&A...550A...1K, 2018ApJ...861...33K} and even subsecond \citep[e.g.,][]{2011SoPh..273..393M, 2019ApJ...872...71Y} time scales, both near the base of the corona and up to 0.7 solar radii above the surface. Although, the wave trains with longer characteristic periods of a few minutes coinciding with EUV observations are also confidently seen in the radio band \citep[e.g.,][]{2009ApJ...697L.108M, 2016A&A...594A..96G, 2017ApJ...844..149K}. In particular, \citet{2013A&A...550A...1K} demonstrated that the local plasma density variations caused by a compressive wave train with the amplitude of a few percent, propagation speed about several hundred km\,s$^{-1}$, and period of a few seconds can explain the observed radio spectra with quasi-periodic fiber bursts. Likewise, using a similar semi-empirical model and the Bayesian and Markov chain Monte Carlo approach, \citet{2018ApJ...861...33K} found a fast wave train modulating the dynamic spectrum of a type III radio burst to propagate at about $657$\,km\,s$^{-1}$, with an oscillation period about 3\,s, and amplitude $<1$\%.

By the observed dynamic properties (i.e., the propagation speed and apparent direction along the local magnetic field, oscillation period, and amplitude) these rapidly propagating quasi-periodic wave trains were associated with fast magnetoacoustic waves guided by certain coronal plasma non-uniformities, such as loops, plumes, current sheets, and magnetic funnels. The quasi-periodic nature of these fast magnetoacoustic wave trains can be attributed either to a periodic driver situated at the coronal base \citep[see, e.g.,][]{2011ApJ...740L..33O, 2018ApJ...860...54O}, or to a self-consistent development from, for example, a flare-caused impulsive perturbation by the waveguide dispersion \citep[see][]{1983Natur.305..688R, 1984ApJ...279..857R, 1994SoPh..151..305M, 1995SoPh..159..399N, 2004MNRAS.349..705N}.
In a driven scenario, the origin of a periodic driver, in turn, requires the interpretation, while the waveguide-caused dispersive effects on fast waves are naturally present. Indeed, fast magnetoacoustic waves guided by plasma waveguides are known to be dispersive, whereas the wave dispersion properties vary with the parameters of the waveguide \citep[see e.g. a series of recent works by][for a comprehensive theoretical treatment of the effects of transverse density profile, magnetic twist, and gravitational stratification of the host plasma structures on the fast wave dispersion]{2015ApJ...810...87L, 2017ApJ...840...26L, 2019MNRAS.488..660L}. The effective formation of dispersively evolving fast wave trains from an aperiodic impulsive driver was modelled by \citet{2013A&A...560A..97P, 2014A&A...568A..20P} for magnetic overdense and underdense funnels; \citet{2014ApJ...788...44M} for dense slabs and current sheets; \citet{2015ApJ...806...56O} and \citet{2015ApJ...814..135S} for cylindrical waveguides.
We need to stress that specific shapes of the transverse density profile of coronal MHD waveguides, for example, a coronal loop, polar plume, jet, streamer, etc., are not known, but are subject to intensive investigation \citep[see, e.g.,][for coronal loops]{2005ApJ...633..499A, 2013ApJ...772L..19B, 2015ApJ...811..104A, 2017A&A...600L...7P, 2017A&A...605A..65G, 2018ApJ...863..167G, 2018ApJ...860...31P}.
In particular, \citet{2017ApJ...847L..21P} demonstrated that the accumulation of nonlinear effects in the dynamics of the fast wave trains trapped in coronal plasma slabs is highly inhibited by strong waveguide dispersion, so that even for the relative perturbation amplitudes of several tens of percent the nonlinear steepening of the wave does not occur. The latter result clearly justifies the applicability of the linear MHD theory to modelling fast wave trains in coronal plasma structures.

Due to the intrinsically non-stationary nature of the dispersively evolving fast wave trains, the wavelet transform was shown to be most suitable for the analysis of their time history. Thus, \citet{2004MNRAS.349..705N} demonstrated that such wave trains have a \lq\lq tadpole\rq\rq-shaped wavelet power spectra, that was extensively used as a characteristic signature for detecting them in observations. Moreover, based on the dependence of the fast wave group speed on the wavenumber along the waveguide axis, \citet{1983Natur.305..688R, 1984ApJ...279..857R} suggested a well developed wave train in a waveguide with sufficiently steep boundaries to have three distinct phases of its time profile, with different behaviours of the instantaneous oscillation period with time. 
In a series of more recent works by \citet{2016ApJ...833...51Y, 2017ApJ...836....1Y} and \citet{2018ApJ...855...53L}, the effect of the waveguide transverse density structuring on the formation and evolution of impulsively excited fast wave trains was investigated, with a particular emphasis on the manifestation of those phases in the obtained time profiles. However, no significant deviations from the \lq\lq tadpole\rq\rq-shaped wavelet power spectrum were found.

In this work, we use recent observations of fast magnetoacoustic wave trains well above the low corona \citep[e.g.,][]{2016A&A...594A..96G, 2018ApJ...861...33K} as a motivation to study characteristic signatures of fully developed fast magnetoacoustic wave trains, i.e. measured sufficiently far from the site of the initial impulsive energy release. We, for the first time, explicitly demonstrate the appearance of three distinct phases of fully developed axisymmetric (sausage) and non-axisymmetric (kink) fast wave trains, which appear in the waveguides with sufficiently steep transverse density profiles and prescribed by a non-monotonic dependence of the fast wave group speed on the parallel wavenumber (Sec.~\ref{sec:dispersion}--\ref{sec:kinks}). The modelling is performed for a dense plasma slab. In Morlet wavelet power spectra, the formation of those distinct phases corresponds to a transition from \lq\lq tadpole\rq\rq-shaped to \lq\lq boomerang\rq\rq-shaped spectral features. The boomerang spectral structures have two well-pronounced arms in the shorter-period and longer-period parts of the spectrum, which could be considered as a new characteristic signature of dispersively evolving fast wave trains to look for in observations (Sec.~\ref{sec:developed}). 
We discuss an observational example identified among the previously published detections of fast wave trains, which demonstrates a clear change of the wavelet spectrum shape from a \lq\lq tadpole\rq\rq\ to a \lq\lq boomerang\rq\rq\ as the wave train propagates through the corona, in full agreement with the results of our modelling (Sec.~\ref{sec:observation}). We outline a roadmap towards the use of these observations and modelling for the purposes of coronal seismology, in particular, for simultaneous probing of the density contrast and steepness of the transverse density profile of the host plasma structures. The discussion of the results and conclusions are given in Sec.~\ref{sec:discussion}.

\section{Waveguide dispersion properties of fast sausage and kink waves} \label{sec:dispersion}

\begin{figure*}
	\centering
	\includegraphics[width=0.495\linewidth]{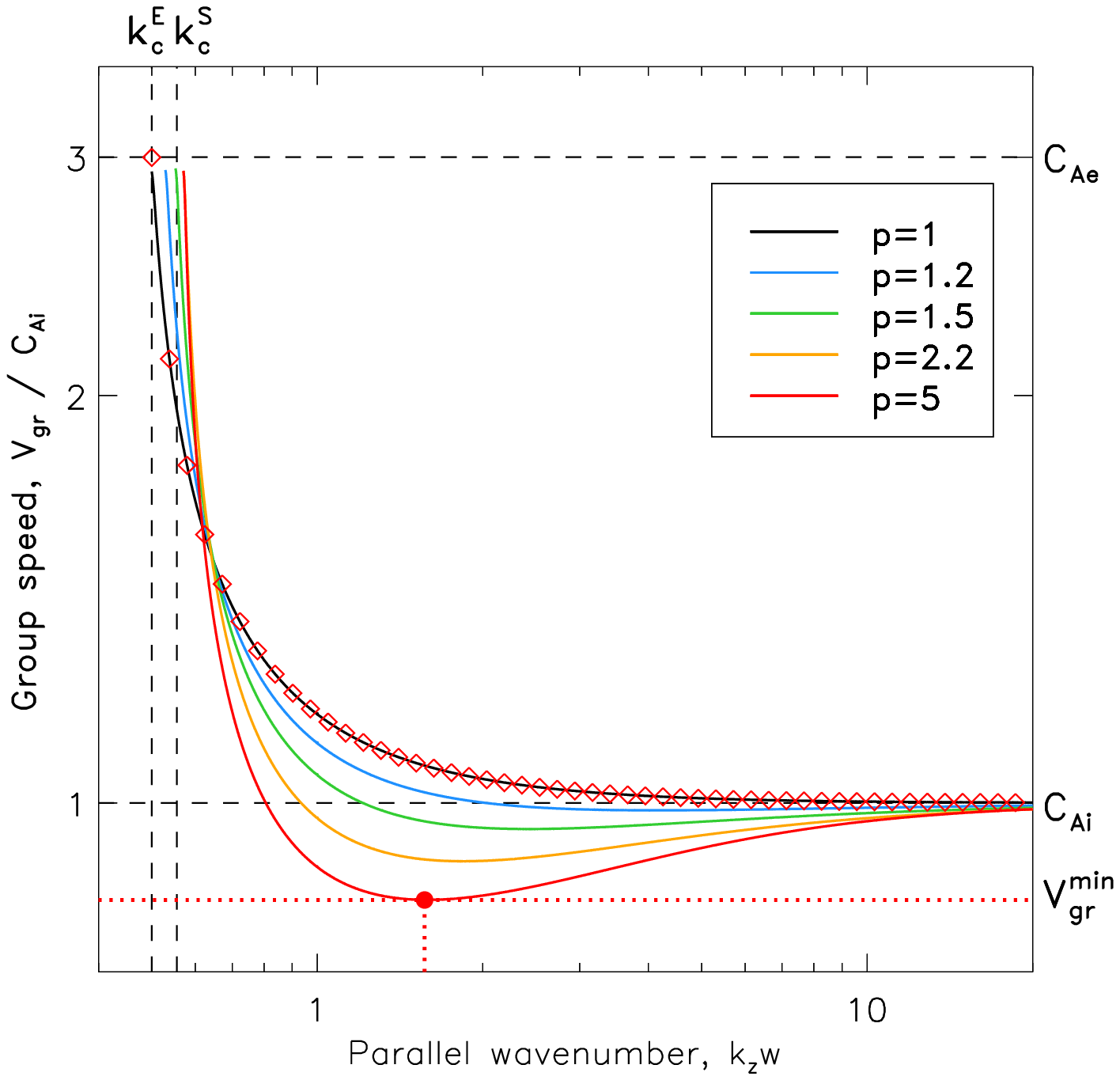}
	\includegraphics[width=0.495\linewidth]{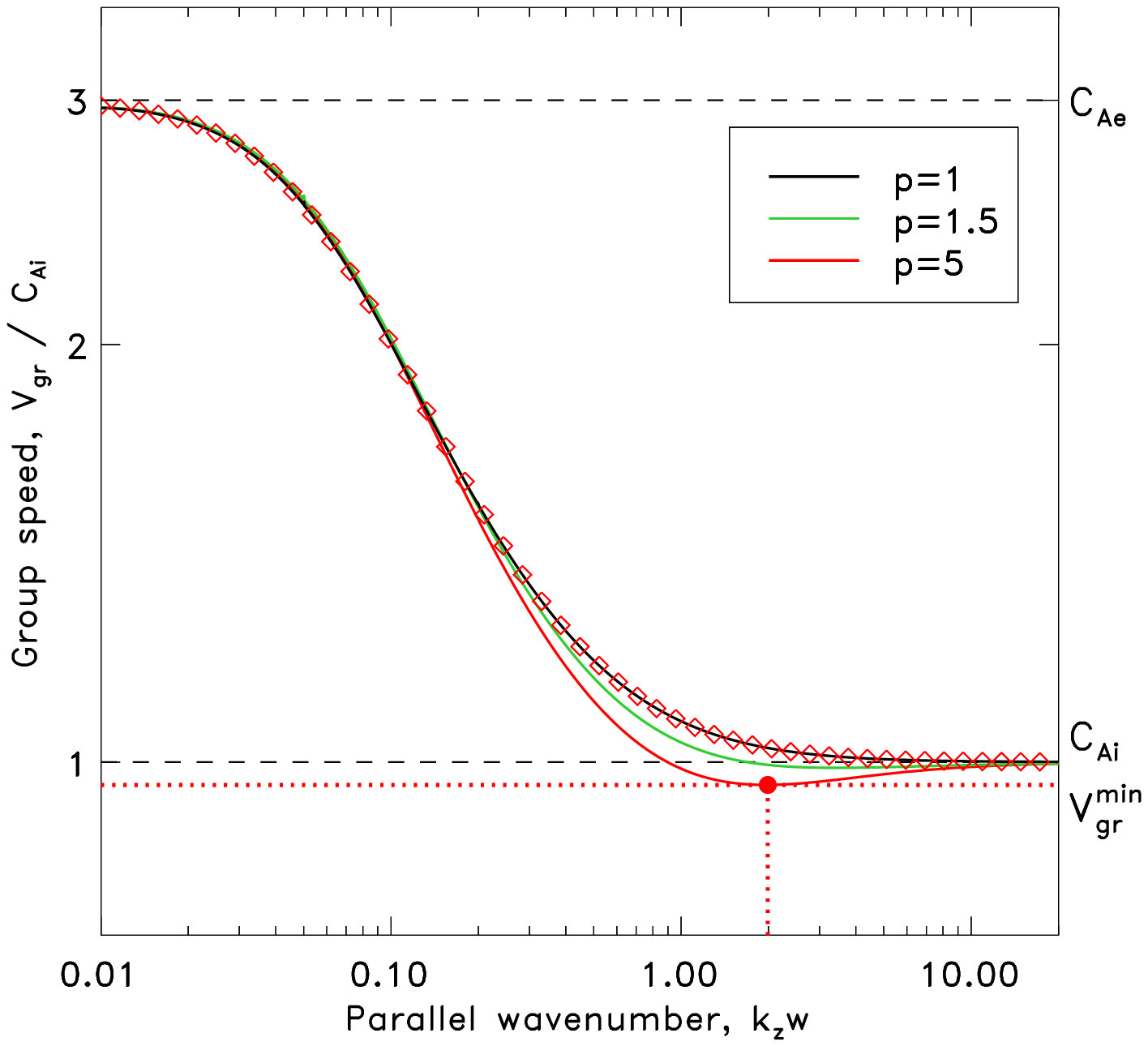}
	\caption{Dependence of the group speed of fast sausage (left) and fast kink (right) magnetoacoustic waves guided by a transverse non-uniformity of plasma density described by Eq.~(\ref{eq:epstein}) with varying steepness parameter $p$ and the Alfv\'en speed ratio $C_\mathrm{Ae}/C_\mathrm{Ai}=3$. %, obtained numerically by the shooting method as described in Sec.~\ref{sec:dispersion}. 
		The red diamonds in both panels show the exact analytic solution for $p=1$ taken from \citet{2003A&A...409..325C}. The sausage cut-off wavenumbers $k_c^\mathrm{E}$ (for a smooth Epstein profile of density with $p=1$) and $k_c^\mathrm{S}$ (for a step-function density profile with $p\gg1$) are given by Eqs.~(\ref{eq:kc_epstein}) and (\ref{eq:kc_step}).
	}
	\label{fig:group_speed}
\end{figure*}

The dynamics of linear fast magnetoacoustic waves guided by a zero-$\beta$ plasma slab stretched along the $z$-axis coinciding with the direction of the magnetic field $B_0$ is described by the following partial differential equation \citep[see, e.g.,][]{2014A&A...567A..24H},
\begin{equation}
C_\mathrm{A}^{-2}(x)\frac{\partial^2 v_x}{\partial t^2}-\frac{\partial^2 v_x}{\partial x^2} + k_z^2v_x=0.
\label{eq:wave_eq}
\end{equation}
Here, $v_x$ is the perturbation of the transverse (across the slab) plasma velocity, $k_z$ is the wavenumber along the $z$ axis, and $C_\mathrm{A}(x)$ is a transverse profile of the Alfv\'en speed determined by the variation of the plasma density across the slab $\rho_0(x)$, as $C_\mathrm{A}(x)=B_0/\sqrt{4\pi \rho_0(x)}$. Equation (\ref{eq:wave_eq}) assumes the plasma to be ideal and uniform along the field and in the $y$ direction. Neither the equilibrium, nor perturbations, depend on $y$. The transverse profile of the plasma density is given by the generalised symmetric Epstein function,
\begin{equation}
\rho_0(x)=(\rho_i-\rho_{e})\sech^2\left(\left[\frac{x}{w}\right]^p\right) + \rho_e,
\label{eq:epstein}
\end{equation}
{\citep[see e.g.][]{1995SoPh..159..399N, 2004MNRAS.349..705N, 2017ApJ...847L..21P}}. In Eq.~(\ref{eq:epstein}), $\rho_i$ is the internal plasma density at $x=0$, $\rho_e$ is the external plasma density at $x\to \infty$, $w$ is the characteristic width of the waveguide, and the parameter $p$ determines the steepness of the density profile that varies from $p=1$ for the Epstein profile to $p>1$ for steeper profiles, tending to a step function considered in \citet{1984ApJ...279..857R} for $p\gg1$.

In addition to the Fourier transform in the $z$ direction already applied to Eq.~(\ref{eq:wave_eq}) and characterised by the wavenumber $k_z$, the Fourier transform in time gives the wave cyclic frequency $\omega$ and allows for separating the variables as $v_x(x,z,t)=U(x)\exp[i(\omega t - k_z z)]$. Thus, Eq.~(\ref{eq:wave_eq}) reduces to the following ordinary differential equation for $U(x)$,
\begin{equation}
\frac{d^2U(x)}{dx^2} + \left[\frac{\omega^2}{C_\mathrm{A}^{2}(x)}-k_z^2\right]U(x)=0.
\label{eq:u(x)}
\end{equation}

In this work, we study dispersive properties of the lowest transverse harmonics of fast magnetoacoustic waves in waveguides with a varying steepness (\ref{eq:epstein}) by the shooting method, i.e. solving Eq.~(\ref{eq:u(x)}) numerically for a fixed value of $k_z$ and searching for a value of $\omega$ that satisfies the boundary conditions $U(x\to\pm\infty)\to0$, by reformulating the boundary problem to an initial value problem. As the initial conditions, we use $U(0)=0$ and $U_x'(0)\ne0$ for axisymmetric (sausage) perturbations, and $U(0)\ne0$ and $U_x'(0)=0$ for non-axisymmetric (kink) perturbations. From the obtained numerical dependencies between $\omega$ and $k_z$ we calculate the wave group speed $V_\mathrm{gr}=d\omega/dk_z$, which is shown in Fig.~\ref{fig:group_speed} for varying values of the waveguide steepness parameter $p$, with the ratio of the external $C_\mathrm{Ae}=B_0/\sqrt{4\pi \rho_e}$ to internal $C_\mathrm{Ai}=B_0/\sqrt{4\pi \rho_i}$ Alfv\'en speeds $C_\mathrm{Ae}/C_\mathrm{Ai}$ set to 3 ($\rho_i/\rho_e=9$). Similar density contrasts have been considered by, e.g., \citet{2016ApJ...823L..16T, 2018A&A...613L...3K, 2019FrASS...6...22P}. This numerical approach was implemented in the computing environment \emph{Maple 2020}. Numerical results were validated by comparison with the exact analytic solution of Eq.~(\ref{eq:u(x)}) existing for a particular case with $p=1$ \citep{1996SoPh..168..273N, 2003A&A...409..325C}, as shown by Fig.~\ref{fig:group_speed}.

The left-hand panel of Fig.~\ref{fig:group_speed} shows the dependence of the group speed $V_\mathrm{gr}$ of guided fast sausage waves on the parallel wavenumber $k_z$ for different values of the steepness parameter $p$. The fast sausage group speed is known to have a cut-off wavenumber $k_c$ discriminating between the trapped ($k_z>k_c$) and leaky ($k_z<k_c$) regimes of the wave dynamics \citep[see, e.g.,][for a comprehensive analysis of the wave leakage in the vicinity of $k_c$]{2014ApJ...781...92V}. In our model, a specific value of the cut-off wavenumber $k_c$ depends on the density ratio ($C_\mathrm{Ae}/C_\mathrm{Ai}$) and steepness ($p$) of the waveguide, and varies between
\begin{align}
&k_c^\mathrm{E} = \frac{1}{w}\sqrt{\frac{2C_\mathrm{Ai}^2}{C_\mathrm{Ae}^2-C_\mathrm{Ai}^2}},&&\mbox{for $p=1$ (smooth profile)},\label{eq:kc_epstein}\\
&k_c^\mathrm{S} = \frac{\pi}{2w}\sqrt{\frac{C_\mathrm{Ai}^2}{C_\mathrm{Ae}^2-C_\mathrm{Ai}^2}},&&\mbox{for $p\gg1$ (step-function)},\label{eq:kc_step}
\end{align}
see \citet{2004MNRAS.349..705N}. The group speed of trapped fast sausage waves is seen to be a monotonic function of $k_z$ for a smooth density profile with $p=1$, and has a well pronounced minimum $V_\mathrm{gr}^\mathrm{min}$ for steeper profiles with $p>1$ \citep[see, e.g.,][]{1995SoPh..159..399N}.
The presence of this dip in the group speed implies that among the ensemble of guided harmonics with different $k_z$, excited by an initial broadband axisymmetric perturbation, will be a relatively narrow (in the $k_z$-space) interval of harmonics propagating at the highest group speeds $C_\mathrm{Ai}<V_\mathrm{gr}<C_\mathrm{Ae}$. These harmonics form a quasi-periodic phase of the whole wave train, and this phase arrives to the observing position first. The wavelengths of these harmonics are the longest among all trapped harmonics. The harmonics travelling at lower speeds $V_\mathrm{gr}^\mathrm{min}<V_\mathrm{gr}<C_\mathrm{Ai}$ will belong to a rather broad interval of $k_z$. Within this interval, the group speed varies non-monotonically with $k_z$, so that there will be pairs of harmonics which have the same group speed but distinctly different parallel wavelengths. These harmonics form a multi-periodic phase of the wave train at some distance from the site of the initial perturbation. The shortest excited harmonic will belong to this phase, provided the driver is sufficiently broadband. At the lowest speed $V_\mathrm{gr}^\mathrm{min}$, there will be a single harmonic trailing behind all other guided harmonics, which can be considered as a periodic phase of the wave train.

For kink waves (right-hand panel of Fig.~\ref{fig:group_speed}), a non-monotonic dependence of the group speed on the parallel wavenumber $k_z$ is also present for steep density profiles with $p>1$. Hence, the formation of those quasi-periodic, multi-periodic, and periodic phases of fast non-axisymmetric wave trains can be expected too. However, in the same waveguide, the dip in the kink group speed is seen to be shallower than that for sausage waves. Moreover, the kink waves are trapped for all $k_z$, which makes the interval of harmonics involved in the quasi-periodic phase broader in the $k_z$-space. A combination of these two effects suggests that in a fully developed fast kink magnetoacoustic wave train, the quasi-periodic phase with the oscillation period gradually decreasing with time will be pronounced stronger than the multi-periodic phase during which the oscillations with distinctly different periods and wavelengths are expected to co-exist.

A similar scenario whereby the initial broadband fast magnetoacoustic perturbation evolves in steep plasma waveguides was discussed and described qualitatively by \citet{1983Natur.305..688R, 1984ApJ...279..857R}. In our work, we present the first quantitative demonstration of the formation of a fast wave train consisting of these three distinct phases.  Characteristic time signatures of those different phases of the fast wave train could be visualised by wavelet analysis, for which we use the \emph{IDL} software package developed by \citet{1998BAMS...79...61T}.

\section{Development of fast sausage wave trains in waveguides with different steepness} \label{sec:development}

Treating a wave train as a linear ensemble of multiple propagating harmonic waves with $\omega$ and $k_z$ prescribed by the waveguide dispersion (Fig.~\ref{fig:group_speed}), a solution to wave equation $(\ref{eq:wave_eq})$ can be written as
\begin{equation}\label{eq:solution}
v_x(x_0,z_0,t)=\sum_{k_z=k_c}^{k_\mathrm{max}}U(x_0,k_{i})A_i\cos(\omega_it-k_iz_0+\phi_i).
\end{equation}
Here, $x_0$ and $z_0$ correspond to an observing position inside the waveguide, $U(x_0,k_{i})$ is determined from a solution of Eq.~(\ref{eq:u(x)}) with $\omega$ obtained from the dispersion relation for each value of $k_z$, and $A_i$ and $\phi_i$ are the amplitudes and initial phases of the individual harmonics $k_i$ taken from the Fourier transform of a Gaussian pulse of an arbitrarily small amplitude, with the root-mean-square width of $0.25w$ and centred at $z=2w$, with which solution (\ref{eq:solution}) is essentially broadband and is localised in the $z$-domain at $t=0$. The highest parallel harmonic $k_\mathrm{max}$ is determined by the full length of the $z$ domain, which was set to $200w$ in this work.
The harmonics with $k_z<k_c$ are not included in solution (\ref{eq:solution}), as in the waveguides with the density contrast
$\rho_i/\rho_e$ lower than 10 they are known to damp by leakage in shorter than one cycle of the oscillation \citep[see, e.g., Eq.~(5) in][and references therein]{2020SSRv..216..136L}. In this work, we determine $k_c$ for various steepness parameters $p$ numerically, as a parallel wavenumber $k_z$ at which the wave group speed $V_\mathrm{gr}$ approaches the external Alfv\'en speed $C_\mathrm{Ae}$ (see Fig.~\ref{fig:group_speed}).

\begin{figure*}
	\centering
	\includegraphics[width=0.495\linewidth]{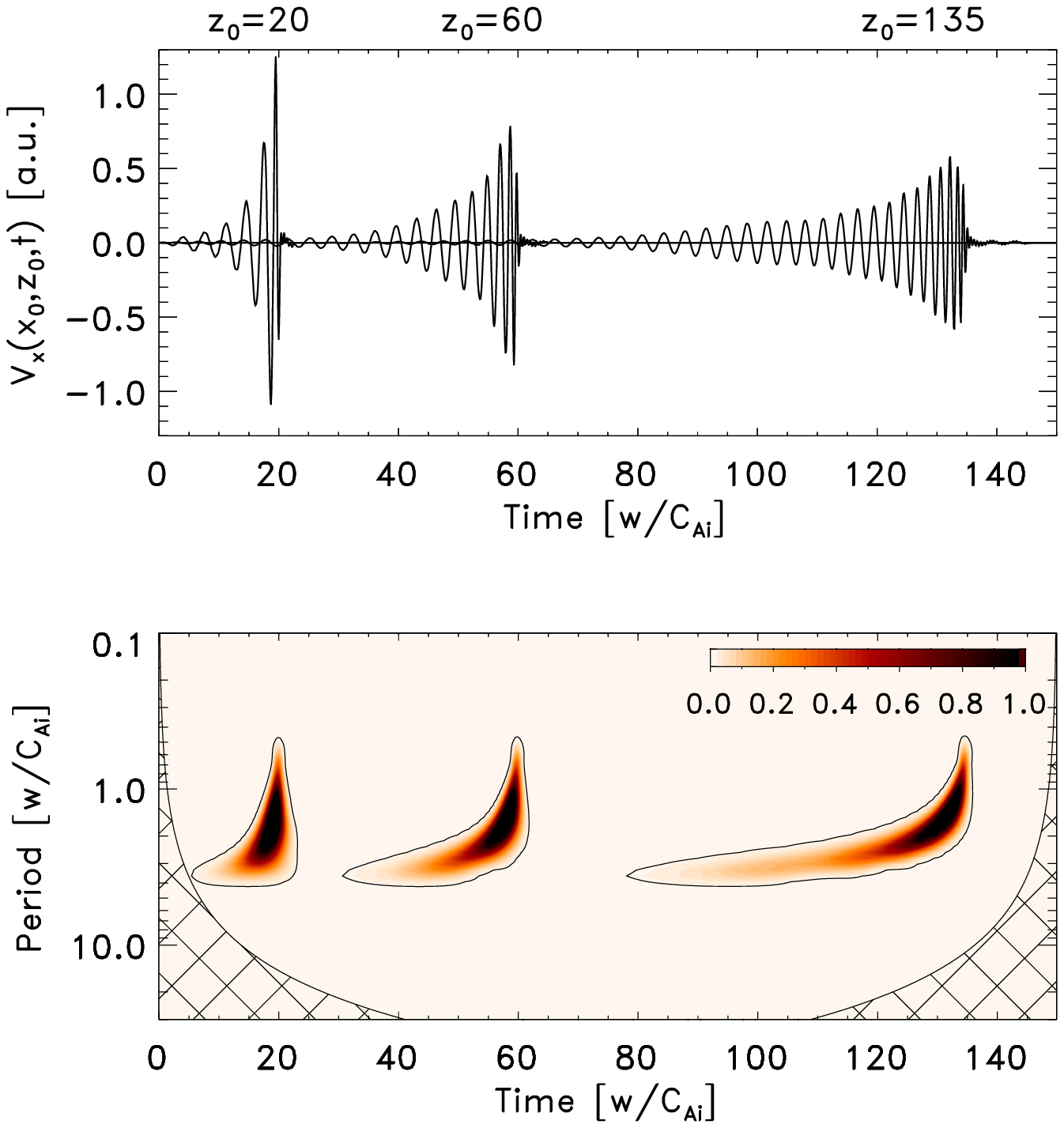}
	\includegraphics[width=0.495\linewidth]{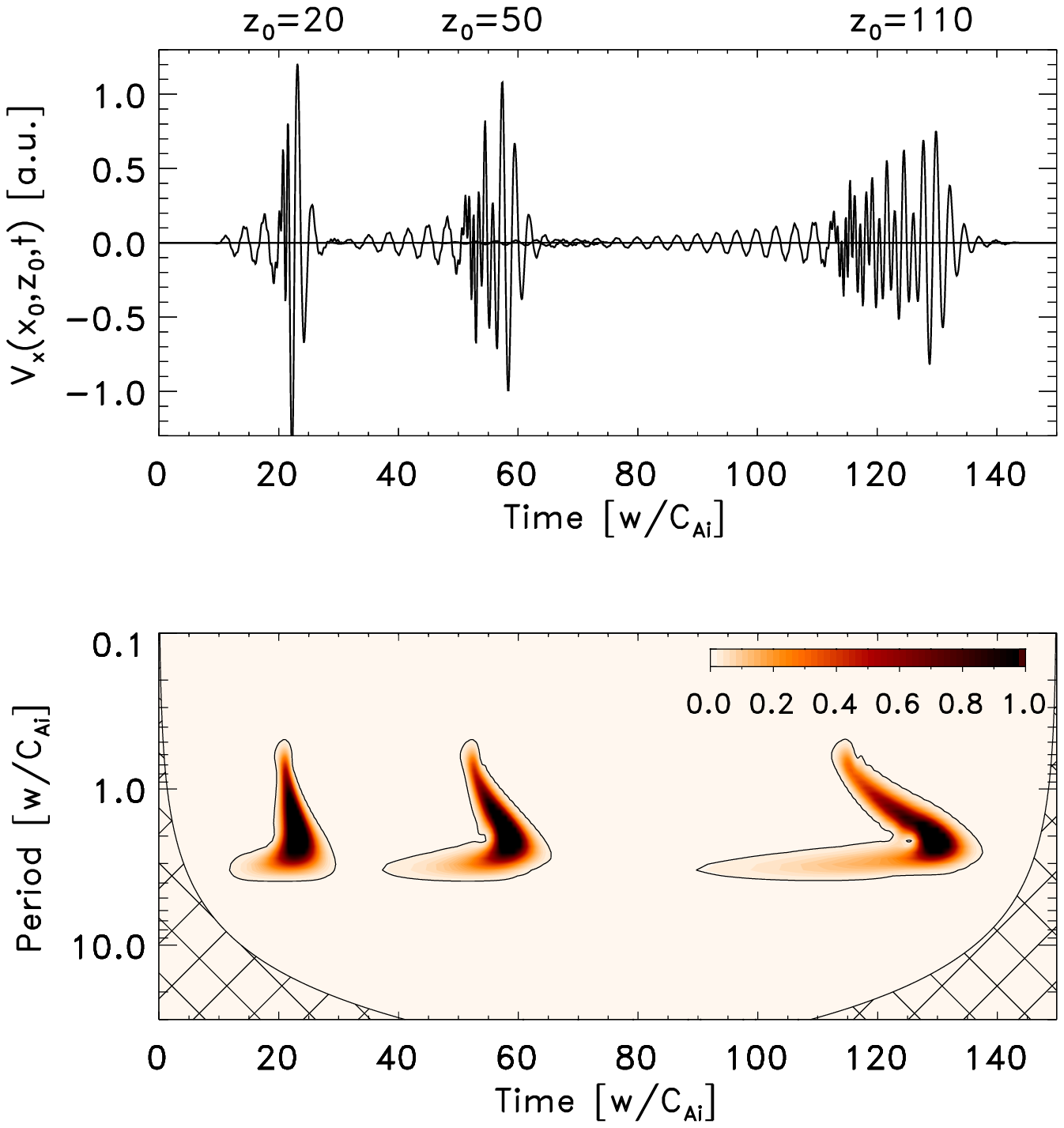}
	\caption{Time profiles (top) and the corresponding Morlet wavelet power spectra (bottom) of fast sausage wave trains determined by Eq.~(\ref{eq:solution}) for $C_\mathrm{Ae}/C_\mathrm{Ai}=3$ and measured at $x_0=0.5w$ and at different parallel distances $z_0$ in either smooth ($p=1$, left) or steep ($p=8$, right) plasma waveguides (\ref{eq:epstein}). The thin solid black lines in the wavelet spectra show the 1\% level of the maximum oscillation power of the corresponding wave train.
	{The colour bars show the wavelet power normalised to its maximum for each individual wave train.}	
	The hatched regions in the bottom panels show the wavelet's cone of influence.
	}
	\label{fig:developemnt_z}
\end{figure*}

Figure~\ref{fig:developemnt_z} demonstrates the time history of a wave train determined by Eq.~(\ref{eq:solution}) and measured at three different distances $z_0$ from the site of the initial perturbation in a smooth waveguide with the density steepness parameter $p=1$ and in a steep waveguide with $p=8$. In a smooth waveguide with $p=1$ (left-hand panels of Fig.~\ref{fig:developemnt_z}), the wave group speed is a monotonic function of the parallel wavenumber $k_z$ as shown by Fig.~\ref{fig:group_speed}, so that each parallel harmonic $k_z$ excited by the initial perturbation propagates at its own speed. This scenario was previously shown to result in the formation of a quasi-periodic wave train with a gradually evolving oscillation period, which is seen as a \lq\lq tadpole\rq\rq\ structure with a narrower-band tail and broader-band head in the Morlet wavelet power spectrum \citep{2004MNRAS.349..705N}.
In the same waveguide, placing the observing position $z_0$ further away from the site of the initial perturbation makes the wave train less compact in space and time, but it still retains a tadpole shape in the wavelet spectrum.

\begin{figure*}
	\centering
	\includegraphics[width=\linewidth]{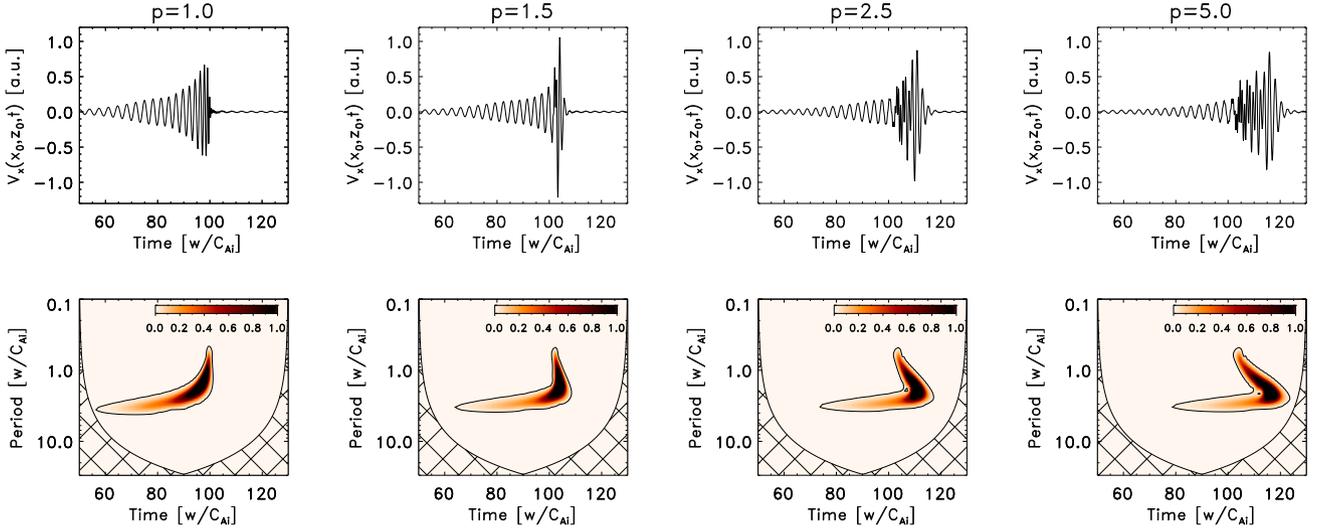}
	\caption{The same as shown in Fig.~\ref{fig:developemnt_z}, but for varying density steepness parameter $p$ and $z_0=100w$.
		%Time profiles (top) and Morlet wavelet power spectra (bottom) of the fast sausage wave trains determined by Eq.~(\ref{eq:solution}) for $C_\mathrm{Ae}/C_\mathrm{Ai}=3$ and measured at $x_0=0.5w$ and $z_0=100w$ in the plasma waveguide (\ref{eq:epstein}) with varying density steepness parameter $p$. The thin solid black lines in the bottom panels show the 1\% level of the maximum wavelet power. The hatched regions in the bottom panels show the wavelet's cone of influence.
	}
	\label{fig:developemnt_p}
\end{figure*}

The evolution of an initial broadband perturbation in a steep waveguide (right-hand panels of Fig.~\ref{fig:developemnt_z}) is seen to be strongly affected by the presence of the dip in the wave group speed (Fig.~\ref{fig:group_speed}).
At $z_0=20w$, the wave trains in smooth and steep waveguides are not much different from each other, both in the time domain and in the wavelet spectrum. However, later stages of the wave train evolution at, for example, $z_0=50w$ and $z_0=110w$, clearly show the formation of distinct phases of the wave train with different time signatures described qualitatively in Sec.~\ref{sec:dispersion}, and the development of a characteristic \lq\lq boomerang\rq\rq\ structure with two well-pronounced arms in the wavelet spectrum of the wave train. A gradual transition from a tadpole-shaped to a boomerang-shaped wave train with increase in the waveguide steepness parameter $p$ from 1 to 5 is shown by Fig.~\ref{fig:developemnt_p}, where all the wave trains are captured at the same observing position $z_0=100w$.

\section{Quasi-periodic, multi-periodic (peloton), and periodic phases of a fully developed fast sausage wave train} \label{sec:developed}

\begin{figure}
	\centering
	\includegraphics[width=\linewidth]{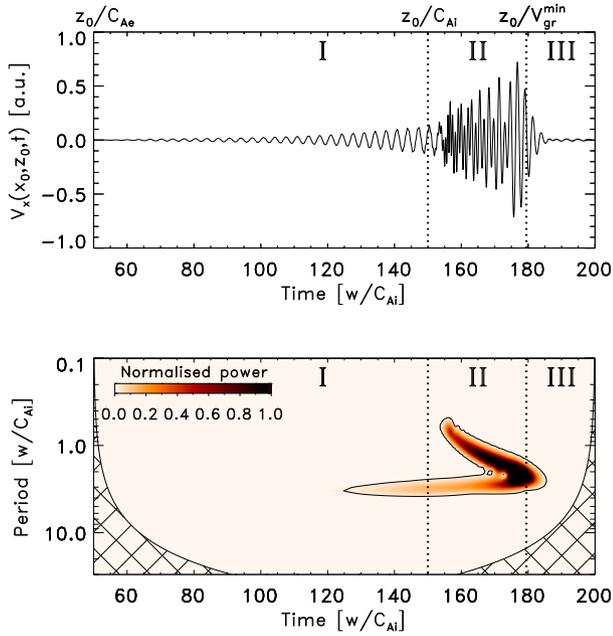}
	\caption{Time profile (top) and Morlet wavelet power spectrum (bottom) of a fully developed fast sausage wave train in a steep plasma waveguide (\ref{eq:epstein}) with $p=8$, obtained from Eq.~(\ref{eq:solution}) for $C_\mathrm{Ae}/C_\mathrm{Ai}=3$, $x_0=0.5w$, and $z_0=150w$. The distinct phases of the wave train are $I$ -- leading quasi-periodic phase; $II$ -- multi-periodic {peloton} phase; $III$ -- periodic Airy phase.
	}
	\label{fig:developed}
\end{figure}

In this section, we give a closer look at a fully developed wave train in a steep waveguide and discuss the distinct phases of its time profile in more detail. Figure~\ref{fig:developed} shows the wave train formed as prescribed by Eq.~(\ref{eq:solution}) in a waveguide with the steepness parameter $p=8$ and measured sufficiently far from the site of the initial perturbation, at $z_0=150w$.

In accordance with the dispersion relation of fast sausage waves in steep waveguides discussed in Sec.~\ref{sec:dispersion} and shown by Fig.~\ref{fig:group_speed}, a relatively narrow interval of harmonics $k_z$ propagating at the group speed monotonically decreasing from $C_\mathrm{Ae}$ to $C_\mathrm{Ai}$ forms a leading (i.e., approaching the observing position $z_0$ in shorter time) phase $I$ of the wave train, lasting from $z_0/C_\mathrm{Ae}$ to $z_0/C_\mathrm{Ai}$ in the time domain.
Due to a small but finite number of harmonics involved in this phase $I$, it could be referred to as \emph{quasi-periodic} (not strictly periodic). In the example shown by Fig.~\ref{fig:developed}, this quasi-periodic phase $I$ is seen as almost monochromatic oscillation both in the time domain and in the wavelet spectrum, because of the effect of the sausage cut-off wavenumber $k_c$ which makes the number of harmonics propagating at $C_\mathrm{Ai}< V_\mathrm{gr} <C_\mathrm{Ae}$ small. The quasi-periodic nature of this phase becomes more apparent for fast kink wave trains, which is demonstrated in Sec.~\ref{sec:kinks}.
This phase is characterised by the longest wavelengths in the wave train. 

The largest number of fast sausage harmonics $k_z$ travelling at $V_\mathrm{gr}^\mathrm{min}< V_\mathrm{gr} <C_\mathrm{Ai}$ constitute another phase of the wave train, indicated as phase $II$ in Fig.~\ref{fig:developed}. It lasts from $z_0/C_\mathrm{Ai}$ to $z_0/V_\mathrm{gr}^\mathrm{min}$ in time and is characterised by the distinctly different parallel harmonics co-existing both spatially and temporally, which is caused by a non-monotonic dependence of the wave group speed on the parallel wavenumber $k_z$ (see Fig.~\ref{fig:group_speed}). In other words, in this interval of $k_z$ there are pairs of longer-wavelength and shorter-wavelength waves excited by the same impulsive driver and propagating at the same group speed ranging from $C_\mathrm{Ai}$ to $V_\mathrm{gr}^\mathrm{min}$. As such, this phase $II$ can be referred to as a \emph{multi-periodic} or \emph{peloton} phase of the wave train, borrowing the terminology used for the main group of riders in road cycling. In the wavelet power spectrum of the wave train, this peloton phase is manifested as two distinct boomerang arms situated at the shorter-period and longer-period parts of the spectrum, which represent those pairs of the parallel harmonics co-existing in this phase of the wave train at each instant of time.

The final short-lived phase of the wave train, phase $III$, is comprised of a single harmonic passing through the observing point $z_0$ at the lowest group speed $V_\mathrm{gr}^\mathrm{min}$. As there is only one harmonic present in this phase, it can be referred to as a \emph{periodic} phase. By analogy with water wave theory, \citet{1983Natur.305..688R,1984ApJ...279..857R} suggested to call this part of the wave train as the Airy phase. In the wavelet spectrum, it corresponds to the elbow of the boomerang, connecting its shorter-period and longer-period arms.

\section{Effect of a dipped group speed on fast kink wave trains} \label{sec:kinks}

Solution (\ref{eq:solution}) can be used for modelling fast kink wave trains too, by applying a non-axisymmetric initial condition for Eq.~(\ref{eq:u(x)}), $U(0)\ne0$ and $U_x'(0)=0$, and setting the minimum value of $k_z$ in Eq.~(\ref{eq:solution}) to zero, as fast kink waves are known to be trapped for all $k_z$ (see Fig.~\ref{fig:group_speed}). All other parameters of the model are kept the same for a clear comparison with the sausage regime.

\begin{figure*}
	\centering
	\includegraphics[width=0.495\linewidth]{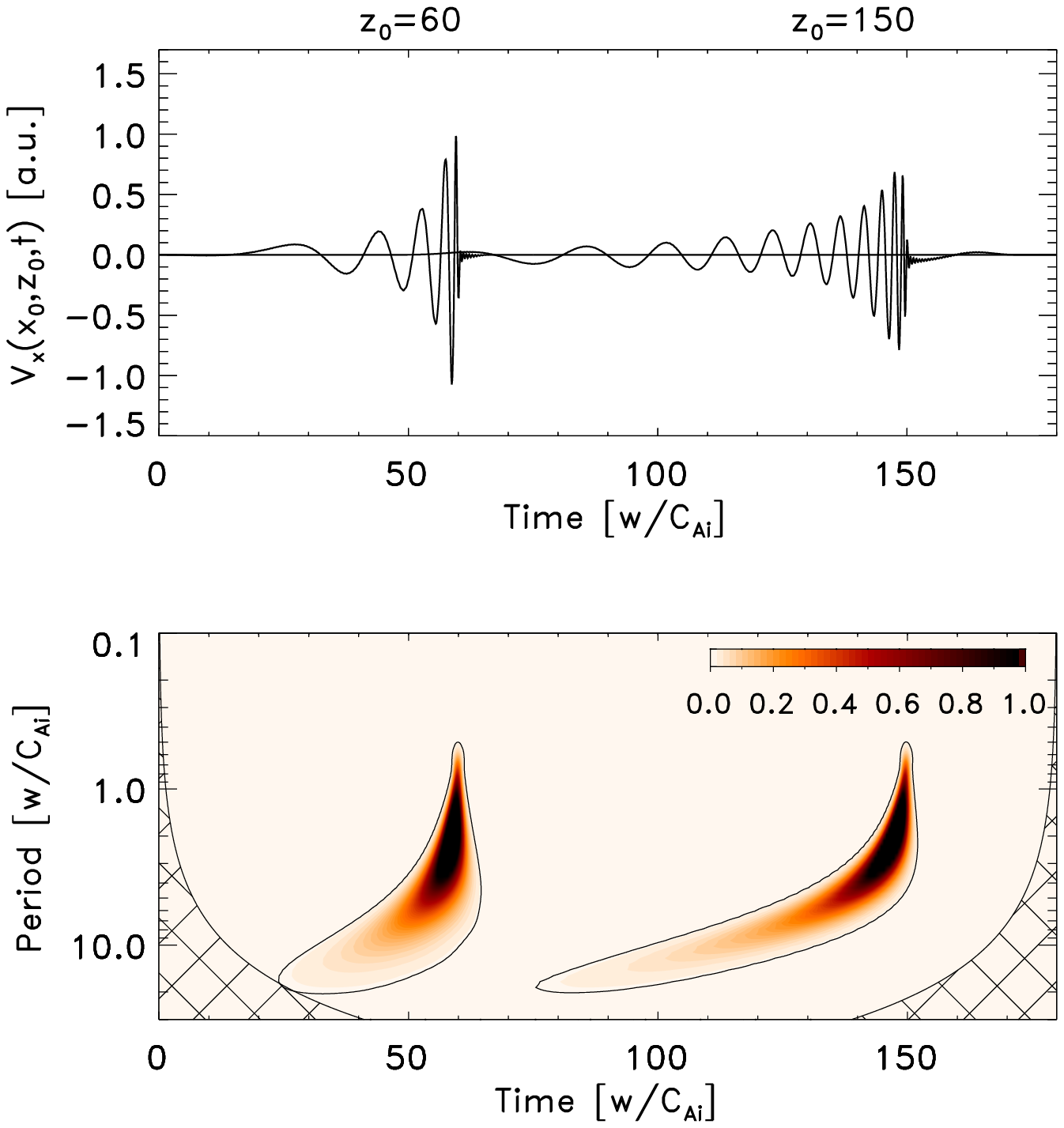}
	\includegraphics[width=0.495\linewidth]{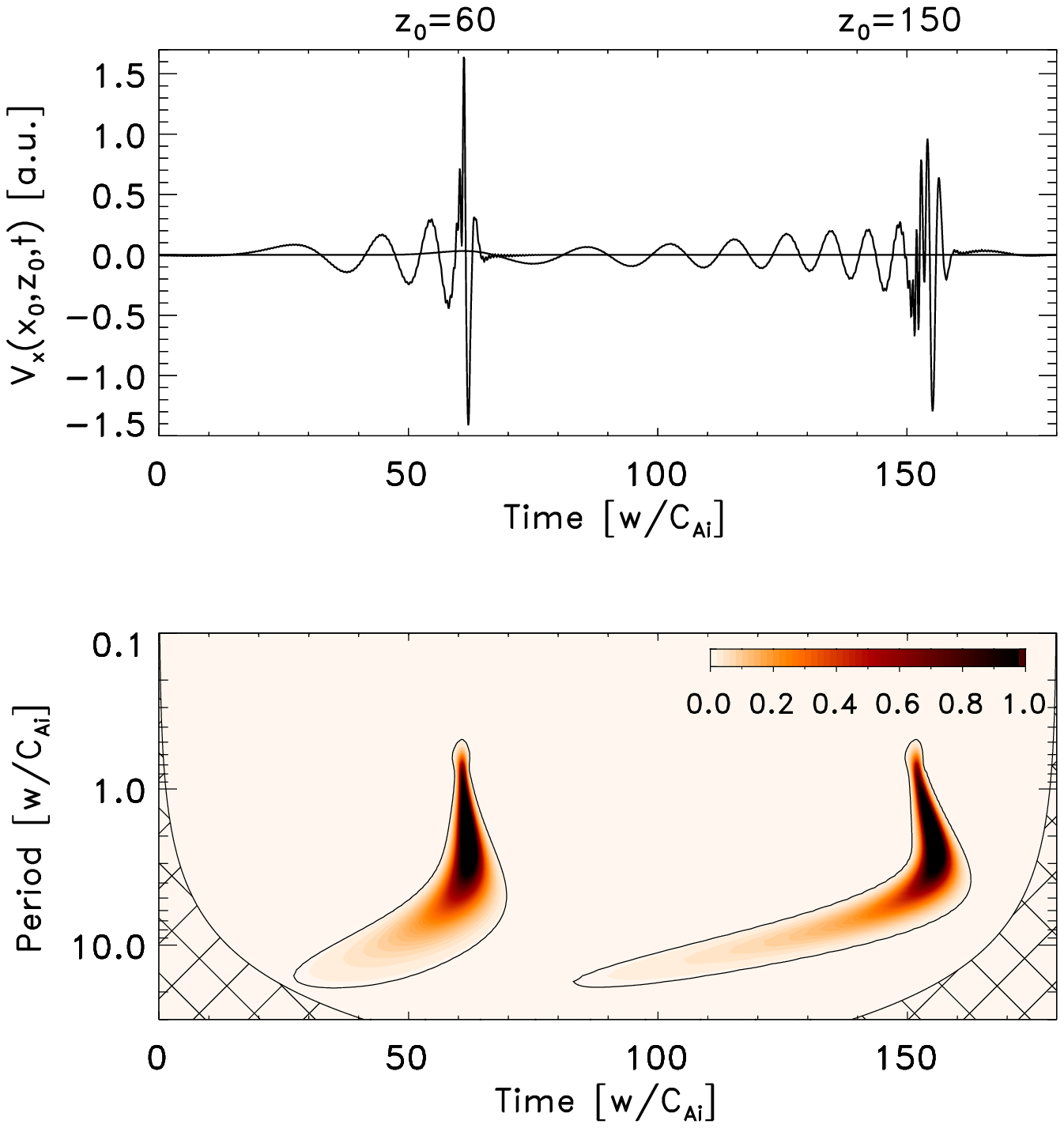}
	\caption{The same as shown in Fig.~\ref{fig:developemnt_z}, but for fast kink wave trains.
	}
	\label{fig:developemnt_kink}
\end{figure*}

The development of fast kink wave trains in a smooth ($p=1$) and steep ($p=8$) plasma waveguide (\ref{eq:epstein}) and their characteristic time signatures in the Morlet wavelet power spectrum are shown in Fig.~\ref{fig:developemnt_kink}. Similarly to sausage wave trains, kink wave trains are seen to change from a tadpole shape to a boomerang shape with the increase in the waveguide steepness parameter $p$, with the quasi-periodic, multi-periodic peloton, and periodic Airy phases clearly present. However, in contrast to the wave trains of a sausage symmetry, the quasi-periodic phase of kink wave trains is seen to be better pronounced with a clear drift of the oscillation period from longer to shorter values with time. This is connected with the absence of the cut-off effect for fast kink waves, that allows the fast-propagating long-wavelength harmonics excited by the initial broadband perturbation to remain inside the waveguide and contribute to the wave train formation. On the other hand, the dip in the group speed of fast kink waves is found to be shallower than that of fast sausage waves in the same waveguide (Fig.~\ref{fig:group_speed}). This makes the peloton phase of a fast kink wave train less developed and pronounced in the time domain and in the wavelet power spectrum. This explicit difference in the development of fast sausage and kink wave trains in the same waveguide can be used for distinguishing between those modes in observations.

\section{Observational illustration and prospects for coronal seismology} \label{sec:observation}

The high temporal resolution traditionally available in the radio band seems to be most suitable for the detection of these boomerang-shaped fast magnetoacoustic wave trains in observations. In particular, signatures of quasi-periodic fast propagating wave trains in solar radio bursts were detected by \citet{2009ApJ...697L.108M} with the characteristic oscillation period about 71\,s; \citet{2011SoPh..273..393M} with multiple periods about 0.5\,s, 2\,s, and 81\,s; \citet{2016A&A...594A..96G} with a period about 1.8\,min; and \citet{2017ApJ...844..149K} with periods ranging from 70\,s to 140\,s. More recently, \citet{2018ApJ...859..154N} observed quasi-periodic pulsations in a solar microflare with periods about 1.4\,s in the radio emission intensity and 0.7\,s in the polarisation signal, which were interpreted in terms of fast sausage oscillations of the flaring loop.  \citet{2018ApJ...861...33K} associated a 3-s quasi-periodic striation observed in the dynamic spectrum of a type III radio burst with the modulation of the local plasma density by a propagating fast wave train. Likewise, \citet{2019ApJ...872...71Y} interpreted a subsecond-period oscillatory dynamics of a radio source in a two-ribbon solar flare by fast propagating magnetoacoustic wave packets in post-reconnection magnetic loops.

An example relevant to this discussion is found in  Fig.~4 of \citet{2011SoPh..273..393M}, which demonstrates wavelet spectra of a fast wave train observed in the decimetric radio emission with fiber bursts by the Ondrejov radio spectrograph after a C-class flare. This wave train was detected at 973--1025~MHz and manifested a frequency drift $\Delta f/\Delta t=8.7$\,MHz\,s$^{-1}$ towards lower frequencies. Assuming the plasma emission mechanism for this observation and taking a hydrostatic density model of the solar atmosphere with the scale height $\Lambda=50$\,Mm (temperature about 1\,MK), this frequency drift can be interpreted as an upward propagation of a wave train at the speed about $870$\,km\,s$^{-1}$, estimated as $(2\Lambda/f)(\Delta f/\Delta t)$ for $f=1000$\,MHz \citep[see e.g.][]{2002SSRv..101....1A}. Interestingly, the shape of the wavelet spectrum of this wave train observed at higher frequencies 1025\,MHz and 1020\,MHz \citep[lower heights, see the bottom panels in Fig.~4 of][]{2011SoPh..273..393M} looks similar to the characteristic shape of a tadpole wavelet structure with a narrower-band tail around 81-s oscillation period and a broader-band head. However, as the wave train propagates upwards \citep[towards lower frequencies 997\,MHz and 973\,MHz, see the top panels in Fig.~4 of][]{2011SoPh..273..393M}, the characteristic shape of its wavelet spectrum clearly changes from a tadpole to a boomerang, with a well pronounced development of the boomerang arm in the shorter-period part of the spectrum, around 30--40\,s \citep[we mention the binary logarithmic scale of the $y$-axes in Fig.~4 of][]{2011SoPh..273..393M}. The described observational properties are fully consistent with the theoretical scenario shown in Fig.~\ref{fig:developemnt_z}. Specifically, it shows the evolution of the wave train shape as it propagates along the waveguide, in full agreement with the theoretical result obtained in this work for a waveguide with a steep transverse density profile. We note that the perturbations of the plasma velocity $v_x$ shown in Fig.~\ref{fig:developemnt_z} can be readily recalculated into the perturbations of the local plasma density using Eq.~(7) of \citet{2003A&A...409..325C}. This allows us to consider Fig.~4 of \citet{2011SoPh..273..393M} as a {possible} observational evidence of the transition from tadpole-shaped to boomerang-shaped fast magnetoacoustic wave trains.

From a seismological perspective, a combination of such an observation {highly suggestive} of a boomerang-shaped wave train with the theoretical model developed in this work offers a unique possibility for probing simultaneously the plasma waveguide depth (Alfv\'en speed ratio $C_\mathrm{Ae}/C_\mathrm{Ai}$) and steepness (parameter $p$). Indeed, according to the model, a quasi-periodic phase of the wave train continues from $z_0/C_\mathrm{Ae}$ to $z_0/C_\mathrm{Ai}$ (see Fig.~\ref{fig:developed}), which allows one to estimate the Alfv\'en speed ratio $C_\mathrm{Ae}/C_\mathrm{Ai}$ from the observed duration of this phase. Likewise, the duration of a peloton phase of the wave train is prescribed to last from $z_0/C_\mathrm{Ai}$ to $z_0/V_\mathrm{gr}^\mathrm{min}$. Having the Alfv\'en speed ratio $C_\mathrm{Ae}/C_\mathrm{Ai}$ estimated at the previous step, its value can be used for solving Eq.~(\ref{eq:u(x)}) parametrically for the steepness parameter $p$ that allows the dip in the fast wave group speed (i.e. ratio between $C_\mathrm{Ai}$ and $V_\mathrm{gr}^\mathrm{min}$, see Fig.~\ref{fig:group_speed}) to be consistent with the observed duration of the peloton phase. As we mentioned earlier in Sec.~\ref{sec:kinks}, the discrimination between the sausage and kink symmetries of the perturbation can be made through the bandwidth of the quasi-periodic phase. The elaboration and practical application of this seismological concept will be a subject of a dedicated follow up work.

%\begin{figure}
%	\centering
%	\includegraphics[width=\linewidth]{obs_example_enhanced.pdf}
%	\caption{Adapted from \citet{2011SoPh..273..393M}.
%		}
%	\label{fig:observation}
%\end{figure}

\section{Discussion and conclusions} \label{sec:discussion}

In this work, we modelled the development of linear dispersively evolving fast magnetoacoustic wave trains in plasma slabs with varying steepness of the transverse density profile. We showed that due to a non-monotonic dependence of the fast wave group speed on the parallel wavenumber in steep waveguides, the initial impulsive perturbation develops into a fast propagating quasi-periodic wave train with three distinct phases: a quasi-periodic phase, a multi-periodic peloton phase, and a periodic Airy phase. This evolution scenario is fully consistent with the wave train signature predicted qualitatively (i.e. without calculations or simulations) by \mbox{\citet{1983Natur.305..688R, 1984ApJ...279..857R}}. These phases form a boomerang structure in the wavelet power spectrum, with two well-pronounced arms in the longer-period and shorter-period parts of the spectrum, that could be considered as a characteristic signature of these wave trains in time-resolved observations. This is in contrast to the wave trains in smooth waveguides, which have no distinct phases in their time history and were previously shown to display tadpoles in the wavelet spectrum. The duration of these phases and how prominent they are in the whole time profile of the wave train, depends on the parameters of the waveguide and the wave perturbation symmetry. In particular, for axisymmetric (sausage) perturbations of the waveguide, the multi-periodic peloton phase was found to be better developed, while the quasi-periodic phase was seen as an almost monochromatic oscillation. For non-axisymmetric (kink) perturbations, the wave trains have a better pronounced quasi-periodic phase with clear decrease in the oscillation period with time, while the multi-periodic peloton phase is also present but rather short-lived and thus less visible.

We have also identified a specific previously published example of a solar coronal fast wave train observed in the radio band, which {strongly suggests} a transition from a tadpole-shaped to a boomerang-shaped wavelet power spectrum, {broadly} consistent with our modelling. The availability of such a {possible} observational confirmation and a theoretically prescribed sensitivity of the distinct phases of the wave train to the parameters of the waveguide open up clear perspectives for using these fast wave trains as a new seismological tool in future. In particular, the ratio of the Alfv\'en speeds inside and outside the waveguide, and steepness of its transverse density profile could be probed simultaneously using high-sensitivity and high-resolution observations of the quasi-periodic and peloton phases of fast wave trains, obtained with the existing (e.g. AIA/SDO, LOFAR) and upcoming (e.g. SKA, METIS/SO, ASPIICS/Proba-3) instruments.

In addition to the parameters of the waveguide and symmetry of the perturbation with respect to the waveguide axis, another important parameter in the fast wave train dynamics is the width of the initial impulsive driver in the $z$-domain, which determines the distribution of the initial energy across the parallel harmonics, i.e., a broadband nature of the perturbation. In this work, this parameter was set to be sufficiently narrow (see Sec.~\ref{sec:development}) to provide the initial perturbation to be essentially broadband and thus to allow for the effective excitation of the parallel harmonics for which the fast wave group speed behaves non-monotonically. For example, \citet{2005SSRv..121..115N} demonstrated that less localised initial perturbations lead to the formation of almost monochromatic fast wave trains, with a poorly pronounced variation of the oscillation period with time. A similar conclusion was drawn in a more recent work by \citet{2017ApJ...836....1Y}, who found that a spatial extent of the initial impulsive driver has to be
comparable to the waveguide width for the effective formation of quasi-periodic fast wave trains. Likewise, \citet{2019A&A...624L...4G} demonstrated that the efficiency of generation of fast wave trains strongly decreases with increasing temporal duration of the impulsive driver. Hence, we expect that the drivers less localised in space and time would lead to a less efficient formation of essentially broadband boomerang-shaped fast wave trains.

A frequency-dependent damping of fast magnetoacoustic waves may also potentially affect the development of the boomerang-shaped wave trains intrinsically comprised of an ensemble of shorter and longer-period harmonics. More specifically, the damping mechanism for sausage oscillations is predominantly associated with the leakage of waves with the wavenumbers below the cut-off value $k_c$ \citep[see e.g.][]{2007AstL...33..706K}.
For the majority of realistic physical conditions used for modelling of fast sausage waves in coronal plasma waveguides, the cut-off wavenumber $k_c$ appears in the long-wavelength part of the spectrum, outside the interval of $k_z$ forming the peloton phase \citep[see][and references therein]{2020SSRv..216..136L}.
Hence, we expect the fast sausage wave leakage to have no significant effect on the formation of the peloton phase. On the other hand, the longer-period quasi-periodic phase of a sausage wave train is strongly affected by this leakage, which makes it apparently narrow-band as we discussed in Sec.~\ref{sec:developed}.
Likewise, the impulsively excited propagating fast kink waves are known to damp in cylindrical waveguides with a smooth density profile due to the phenomenon of resonant absorption, with the damping time/length proportional to the oscillation period \citep[see e.g.][]{2011SSRv..158..289G, 2013A&A...551A..39H, 2013A&A...551A..40P}. This may lead to the suppression of shorter scales in the peloton phase of a fast kink wave train. However, as we demonstrated in Sec.~\ref{sec:kinks}, the fast kink wave trains are dominated by the longer-period quasi-periodic phase for which the effect of resonant absorption is weaker. Nonetheless, the question of a relative efficiency of the processes of frequency-dependent damping and formation of boomerang-shaped fast wave trains remains open and requires a dedicated parametric study.

Another potentially interesting direction for the development of this work is to extend it upon the waveguides with other transverse density profiles. For example, \citet{2016ApJ...833...51Y} demonstrated that fast wave group speed can have multiple extrema for sufficiently wide linear transition layers sandwiched between internal (with a uniform density $\rho_i$) and external (with a uniform density $\rho_e$) regions. Clearly, the existence of multiple dips in the fast wave group speed would make the time evolution of the initial impulsive perturbation even more complicated than described in this work, potentially with more distinct phases and more sophisticated wavelet structures of the wave train. Similarly to the results of this work, revealing these characteristic signatures could be used as a seismological indicator of the transverse structuring of a hosting plasma waveguide.

\section*{Acknowledgements}
D.Y.K. and V.M.N. acknowledge support from the STFC consolidated grant ST/T000252/1. D.Y.K. was supported by the Ministry of Science and Higher Education of the Russian Federation. V.M.N. acknowledges the Russian Foundation for Basic Research grant No. 18-29-21016.

\section*{Data availability}
The data underlying this article are available in the article and in the references therein.

\bibliographystyle{mnras}
%\bibliography{boomerangs_kolotkov} % if your bibtex file is called example.bib

\bsp	% typesetting comment
\label{lastpage}
\end{document}